\documentclass[prd,aps,12pt]{revtex4}
\usepackage{latexsym}
\usepackage{amssymb}
\usepackage[dvips]{graphicx}
\begin{document}
\newcommand{\beq}{\begin{equation}}
\newcommand{\eeq}{\end{equation}}
\newcommand{\ove}{\overline}
\newcommand{\half}{\frac 1 2 }
\newcommand{\fourth}{\frac 1 4}
\newcommand{\Fstar}{\raisebox{.2ex}{$\stackrel{*}{F}$}{}}
\newcommand{\Pstar}{\raisebox{.2ex}{$\stackrel{*}{P}$}{}}
%
\newcommand{\et}{{\em et al}}
\newcommand{\ie}{{\em i.e.$\;$}}
\newcommand{\call}{{\cal L}}
%
%
\newcommand{\Prd}{Phys.  Rev. D$\;$}
\newcommand{\Prl}{Phys.  Rev.  Lett.}
\newcommand{\Plb}{Phys.  Lett.  B}
\newcommand{\Cqg}{Class.  Quantum Grav.}
\newcommand{\Np}{Nuc.  Phys.}
\newcommand{\Grg}{Gen.  Rel.  \& Grav.$\;$}
\newcommand{\Fp}{Fortschr.  Phys.}
\newcommand{\Sch}{Schwarszchild$\:$}
\renewcommand{\baselinestretch}{1.2}

\title{Static and spherically symmetric black holes in $f(R)$ theories}
\author{Santiago Esteban Perez Bergliaffa and Yves Eduardo Chifarelli de Oliveira Nunes}
\affiliation{Departamento de F\'{\i}sica Te\'{o}rica,
Instituto de F\'{\i}sica, Universidade do Estado de Rio de
Janeiro, CEP 20550-013, Rio de Janeiro, Brazil}

\date{\today}
\vspace{.5cm}

\begin{abstract}
We study some features of static and spherically symmetric solutions (SSS) with a horizon in $f(R)$ theories of gravitation by means of a near-horizon analysis. A necessary condition for an $f(R)$ theory to have this type of solution is obtained. General features of the effective potential are deduced, and it is shown that there exists a limit on the curvature at the horizon,
in both cases for any $f(R$). Finally, we calculate the expression for the energy of the collision of two masive particles in the center of mass frame. 
\end{abstract}


\vskip2pc
\maketitle

\section{Introduction}

Although gravity has been shown to be with high accuracy in accordance with General Relativity (GR)
in a number of situations in which the curvature is small \cite{will}, 
there is no observational evidence of the behaviour
of the gravitational field
for very large values of the curvature. In this regard,
objects such as black holes and neutron stars are the ideal places to 
look for deviations from GR in the strong regime \cite{psaltis}.
The task of understanding what kind of deviations can be expected, and their relation to observable quantities
is of relevance both theoretically and from the observational point of view. The latter is important in view
of developments 
that 
offer the prospect of surveying phenomena 
occuring in the vicinity of the horizon in the near future, such as 
the possibility of obtaining ``black hole images" \cite{loeb}
\footnote{Another phenomenon which probes
the strong-field regime, but which is not observable with present-day technology,
is the strong gravitational lensing \cite{virb}.}.

As a first step in discussing deviations from General Relativity in 
strong-field gravity, we will examine here  
some features of 
static and spherically symmetric (SSS) black hole solutions
in theories with a Lagrangian that is a function of the Ricci scalar $R$ \footnote{These theories have been analyzed in detail lately, due to the fact that they can describe the 
observed accelerated expansion of the universe (which takees place in the low-curvature regime)
without the introduuction of  
dark energy \cite{fara}.}. Different aspects of this type of solutions in $f(R)$ theories have been previously in 
discussed in \cite{sss}, mostly resorting to exact solutions and/or phase space analysis. 
We will take here a complementary path which consists in extracting information about relevant quantities from the behaviour of the geometry near the horizon. For an arbitrary $f(R)$,
it will be assumed 
that a SSS black hole solution 
exists, being described by 
a general metric adapted to these symmetries. 
As a result of expanding the metric functions in series of the distance to the horizon (whose radius
$r_0$ is given by
$g_{00}(r_0)=0$), and using the equations of motion for the metric, we shall obtain 
a  
necessary condition that the $f(R)$ must satisfy for the existence of the SSS black hole
solution. It will also be shown that the 
near-horizon geometry is constrained by the EOM, and the consequences of these constraints in the redshift, the curvature, and the energy of a collision of particles in the center of mass frame will be analyzed.
Let us begin by presenting the relevant equations in the next section.

\section{Equations of motion at the horizon}
\label{eq}
The vacuum equations of motion for an $f(R)$ theory are given by
\begin{equation}  \label{fieldeq}
\frac{df}{dR}R_{\mu\nu} - \frac{f}{2}g_{\mu\nu} - \left(\nabla_{\mu}\nabla_{\nu}-
g_{\mu\nu}\,\Box \,\right)\frac{df}{dR} = 0,
\end{equation}
along with the trace of Eq.(\ref{fieldeq}):
\begin{equation}  \label{trace_eq}
3\Box \frac{df}{dR} + \frac{df}{dR} R - 2f = 0.
\end{equation}
In the case of a SSS metric
in Schwarzschild's coordinates,
the nonzero equations following from Eqn.(\ref{fieldeq}) are
\beq
\frac{df}{dR}R_{00}+g_{00}A=0,
\label{00}
\eeq
\beq
\frac{df}{dR}R_{11}+g_{11}A-\frac{d^{\:3}f}{dR^{\:3}}(\partial_1 R)^2-\frac{d^{\:2}f}{dR^{\:2}}\nabla_1(\partial_1R)=0,
\label{11}
\eeq
\beq
\frac{df}{dR}R_{22}+g_{22}A=0,
\label{22}
\eeq
where
\beq
A\equiv -\frac f 2 +\frac{d^{\:3}f}{dR^{\:3}}g^{11}(\partial_1 R)(\partial_1 R)+\frac{d^{\:2}f}{dR^{\:2}}\Box R .
\eeq
To describe the static and spherically-symmetric spacetime the metric 
\beq
ds^2=-e^{-2\phi (r)}\left(1-\frac{b(r)}{r}\right)dt^2+\frac{dr^2}{1-\frac{b(r)}{r}}+r^2d\Omega^2
\label{metric}
\eeq
will be adopted, where the function $\phi$ is known as the anomalous redshift. 
We shall assume that there is a horizon at $r=r_0$ \footnote{In case of multiple horizons,
$r_0$ designates the radius of the most external one.}, 
where $r_0$ is given implicitly by
$b(r_0) = r_0$. From this expression we see that  
the dependence of $r_0$ with the mass 
will likely be different from the linear relation 
$r_0=2M$ in Schwarzschild's solution.
We assume in the following that all the functions in the metric, as well as the derivatives 
of $f(R)$, 
can be developed in series
around the horizon, so that
$$
b(r)=b_0+b'_0(r-r_0)+\half b''_0(r-r_0)^2+...
$$
$$
\phi (r) = \phi_0 + \phi '_0 (r-r_0) + \half \phi ''_0 (r-r_0)^2+...,
$$
$$
\frac{df}{dR}=\left.\frac{df}{dR}\right|_0 + \left[\frac{d}{dr}\left(\frac{df}{dR}\right)\right]_0(r-r_0) + ...,
$$
where the subindex zero indicates that the corresponding quantity is evaluated
at the horizon, and the prime denotes derivative with respect to the coordinate $r$. Replacing these expressions in the EOM and taking the limit
$r\rightarrow r_0$ is a rather long and straightforward calculation that requires care, because
there will be both finite terms and terms that diverge as $(r-r_0)^{-1}$
in this limit. The latter arise from the ``11'' component 
of the Ricci tensor, given by
$$
R_{11}=-\frac{1}{2r^2\left(1-\frac{b}{r}\right)}\left\{2r^2\left(1-\frac{b}{r}\right)(\phi ''-\phi '^2)+3\phi '(b-rb')+b''r\right\}
$$
and also from the term $\Box R=g^{11}\nabla_1\partial_1R$. 
For the EOM to be satisfied at the horizon, we
must impose that both the finite and the divergent terms be zero. Doing so leads to the following 
relations: from Eqn.(\ref{00}) we obtain
\beq
3\phi'_0r_0(1-b'_0)+b''_0r_0-2b'_0=0.
\label{c1}
\eeq
The finite part of Eqn.(\ref{11}) yields 
\beq
\left.\frac{df}{dR}\right|_0\left[3\phi'_0b'_0-b''_0+2\phi'_0+2b'_0(r_0\phi''_0-r_0\phi_0^{'2}-\phi'_0)+4\frac{b'_0}{r_0}\right]
+2b'_0R'_0\left.\frac{d^{\:2}f}{dR^{\:2}}\right|_0=0,
\label{c2}
\eeq
with
$$
R'_0=-\frac{1}{r_0^2}\left\{b''_0(1-3\phi'_0r_0)+(1-b'_0)(5r_0\phi''_0-2r_0\phi_0^2-2\phi'_0)
+r_0b'''_0-4\frac{b'_0}{r_0}\right\},
$$
while the divergent part yields
\beq
2\left.\frac{df}{dR}\right|_0b'_0+r_0^2f_0=0.
\label{c3}
\eeq
From the finite part of Eqn.(\ref{22}) we get
\beq
-2\left.\frac{df}{dR}\right|_0\phi'_0+r_0R_0^{'2}\left.\frac{d^{\:3}f}{dR^{\:3}}\right|_0+r_0B_0
\left.\frac{d^{\:2}f}{dR^{\:2}}\right|_0=0
\label{c4}
\eeq
where $B_0=B_0(b'_0,b''_0,b'''_0,b^{iv}_0,\phi'_0,\phi''_0,\phi'''_0)$ is a rather long expression 
which we shall not use in this paper. 
Finally, the divergent part of Eqn.(\ref{22}) gives
\beq
-5r_0^2\phi_0''(b'_0-1)+r_0\phi'_0(-3r_0b''_0+2(b'_0-1))+2r_0^2\phi_0^{'2}(b'_0-1)+
r_0(b'''_0r_0+b''_0)-4b'_0=0.
\label{fi2}
\eeq
which is equivalent to 
$ 
R\:'_0=0.
$
Assuming $\left.\frac{df}{dR}\right|_0\neq 0$ (we shall see below that this is a reasonable asssumption), it follows from Eqn.(\ref{c2})
that \beq
3\phi'_0b'_0-b''_0+2\phi'_0+2b'_0(r_0\phi''_0-r_0\phi_0^{'2}-\phi'_0)+4\frac{b'_0}{r_0}=0.
\label{c22}
\eeq
It can also be shown from these relations that the trace equation at $r=r_0$, given by
$$
\left. \frac{df}{dR}\right|_0R_0-2f_0+\frac{3}{2r_0}(1-b'_0)\left. \frac{d^{\:2}f}{dR^{\:2}}\right|_0 R\:'_0=0
$$
is identically zero.
These equations and some of their consequences will be analyzed in the following. 
Before closing this section, let us remark that although these relations were obtained using
a series development around $r=r_0$, they are exact, in the sense that the higher order terms 
go to zero when $r \rightarrow r_0$.

\section{Near-horizon behaviour}

It will be shown in this section that
information about the near-horizon geometry and the $f(R)$ can be extracted
from Eqns.(\ref{c1}), (\ref{c3}), (\ref{fi2}) and (\ref{c22}), 
and used to study relevant quantities.
To begin with, 
taking into account that 
$R_0=-4b'_0/r_0^2$, Eqn.(\ref{c3}) can be rewritten as 
\beq
\frac{f_0}{\left.\frac{df}{dR}\right|_0}=\frac{R_0}{2}.
\label{nc1}
\eeq
which furnishes an easy-to-use, coordinate-independent necessary condition for a given $f(R)$ to have SSS black hole solutions.
In particular, it follows that Schwarzschild's metric (for which $b(r)=$constant)
is not a solution of those theories for which 
$$
\frac{f_0}{\left.\frac{df}{dR}\right|_0}\neq 0,
$$
a result which agrees with the conclusions obtained in 
\cite{psaltis2} and \cite{dunsby2}. This expression can be used to test 
whether a given theory has SSS black hole solutions. For instance, it follows from
Eqn.(\ref{nc1}) that
the theory defined by $f(R)=\alpha R^n$
may only have SSS black hole solutions for $n=2$.\\
This condition may be strengthened by the use of the inequality $\frac{df}{dR} >0$, which must be satisfied in order to avoid ghost-like behaviour of cosmological perturbations \cite{defe}. Using the latter condition along with Eqn.(\ref{nc1}), we conclude that for a given $f(R)$ theory to be free of ghost-like cosmological perturbations and to have a SSS solution with $R_0>0$ ($R_0<0$),
$f_0$ must be positive (negative). In particular, in the case of 
Schwarzschild's spacetime, the condition $f_0=0$ must be met.
 
Going back to the set of equations obtained in the previous section, 
notice that Eqns.(\ref{c3}) and (\ref{c4}) involve
the function $f$ and its derivatives at the horizon, while Eqns.(\ref{c1}), (\ref{fi2}) and
(\ref{c22})
are constraints
on the geometry in the neighbourhood of $r_0$. In particular, from Eqns.(\ref{c3}) and
(\ref{c22}) the first and second derivatives of $\phi$ at the horizon can be expressed as functions of $r_0$, $b'_0$, and $b''_0$. Hence, due to the equations of motion, the near-horizon geometry
up to second order in the distance to the horizon is determined by the function $b$. This will be exemplified below in the case of the effective potential for photons.
   
Another condition on the near-horizon metric comes from the redshift, given by
$$
1+z = \frac{\nu_R}{\nu_E} = \left[\frac{g_{00}(E)}{g_{00}(R)}\right]^{1/2}.
$$
We shall assume here that the reception point $R$ is at infinity, and the emission point $E$ is near the horizon. It follows that
$$
1+z \approx \left[\frac{e^{-2\phi_0}}{r_0}(1-b'_0)(r-r_0)\right]^{-1/2}
$$
sufficiently near the horizon. Hence, the condition
\beq
1-b'_0 > 0 
\label{cr2}
\eeq
must be satisfied if the redshift is to be well-defined near the horizon.
In fact, this is the condition for the metric to have the right sign near the horizon, and also for the tidal forces at the horizon be coincident in sign (and finite) with 
those of Schwarzschild's black hole (see the Appendix). 
Using the relation $R_0=-4b'_0/r_0^2$, this condition entails the existence of a limit 
for the curvature at the horizon:
$$
R_0 > -\frac{4}{r_0^2}.
$$
Inequality (\ref{cr2}) is also important for the motion of particles near the horizon. In the case of massless particles, the effective potential is 
defined by
$$
V_{{\rm eff}}= \frac{L^2}{r^3}e^{-2\phi }(r-b).
$$
With the definition $v=V/L^2$, it follows that the first derivative at the horizon is given by
\beq
\left.\frac{dv}{dr}\right|_0 = \frac{1}{r_0^3}e^{-2\phi_0}(1-b'_0).
\label{ep1}
\eeq
Hence, due to Eqn.(\ref{cr2}), the first derivative of the effective potential is positive, as in 
the case of Schwarzschild's solution. Since at infinity the effective potential must go to zero, the solution must have at least one unstable circular orbit for photons. 
Qualitative differences in the effective potential appear only to second order 
in the distance to the horizon, with the 
second derivative given by
$$
\left.\frac{d^2v}{dr^2}\right|_0 =-\frac{1}{r_0^4}e^{-2\phi_0}\left[2(1-b'_0)(3+2\phi'_0r_0)+r_0b''_0\right].
$$
Using Eqn.(\ref{c1}), we can eliminate $\phi'_0$, yielding 
\beq
\left.\frac{d^2v}{dr^2}\right|_0 =-\frac{1}{r_0^4}e^{-2\phi_0}\left[6-\frac{10}{3}b'_0-\frac{1}{3}b''_0r_0\right],
\label{ep2}
\eeq
in such a way that $\left.\frac{d^2v}{dr^2}\right|_0$ depends of $b_0''$, on which we have no constraints \footnote{Going to third order, the second derivative of $\phi$ evaluated at the horizon that would appear can be expressed in terms of the derivatives of $b$ at the horizon using Eqn.(\ref{fi2}).}. Notice also that $\phi_0$ acts as a scale in the series development
near the horizon.
 
\subsection{Collisions}

Another phenomenon for which there may be differences between the type of black hole under study here and Schwarzschild's is the collision of two particles. As shown in \cite{col1}, the maximal collision energy in the center of mass system for two particles of mass $m$ moving in Schwarszchild's geometry , given by $E_{CM}(r_0)=2\sqrt 5 m$,
is reached at the horizon, and is attained when the two particles have angular momentum equal in magnitude and opposite in sign. Let us see how this result changes for the case at hand.\\
The energy in the center of mass system is given by
\cite{banados}  
$$
E_{CM} = \sqrt{2}m\sqrt{1-g_{\mu\nu}v^\mu_{(1)}v^\nu_{(2)}},
$$
where $v^\mu_{(i)}$ is the 4-velocity of each particle. In turn, the 4-velocities are furnished by the geodesic equations, which in the case of the metric (\ref{metric}) are:
$$
\frac{dr}{d\tau} = -\sqrt{e^{2\phi}-\left(1-\frac{b}{r}\right)\left(1+\frac{\tilde L^2}{r^2}\right)},
$$
$$
\frac{d\phi}{d\tau} = \frac{\tilde L}{r^2},
\;\;\;\;\;\;\;\;\;\;\;\;\;\;\;\;
\frac{dt}{d\tau} = \frac{e^{2\phi}}{1-b/r}, 
$$
where $\tilde L$ is the angular momentum per unit mass, we are taking $\theta = \pi/2$ and assuming that the particles start from rest at infinity. 
Using these equations and the metric given in Eqn.(\ref{metric}) in the expression for $E_{CM}$,
a straightforward calculation shows that at the horizon,
$$
E_{CM}(r_0) = m\sqrt{4+\frac{(\tilde L_1-\tilde L_2)^2}{b_0^2}},
$$
where $\tilde L_i$ is the angular momentum of each particle. This expression reduces to the one quoted above in the case of Schwarszchild, $b_0=r_0=2$ (with $M=1$), and $\tilde L_1=-\tilde L_2=4$ \footnote{These are the maximum allowed values such that
particles reach the horizon with maximum tangential velocity \cite{col1}.}. 
We see that in general $E_{CM}$ will be different $E_{CM}(r_0)=2\sqrt 5 m$, due to the 
fact that the relation that determines the radius of the horizon ($b(r_0) = r_0$)
is different from $r_0 = 2$.

\section{Discussion}
\label{di}

As a first step in the investigation of strong gravity efffects in $f(R)$ theories, we have studied the near-horizon behaviour of a static and spherically symetric black hole solution. 
We have obtained a necessary condition for a given $f(R)$ to have such a solution, and 
showed that the equations of motion constrain the near-horizon geometry. 
These constraints entail that there is a maximum allowed value for the curvature at the horizon,
and also that the effective potential differs qualitatively from that of Schwarzschild's
only at second order in the distance to the horizon. In particular, it was concluded that
there must be an unstable orbit for photons. We have also obtained the expression for the center of mass energy
for the collision of two particles of mass $m$  at the horizon, which can depends on the function $b(r)$ of the metric, evaluated at $r_0$. Hence, this energy is different from the analog expression in GR. 
All these results were obtained by means of a local analysis, without any constraint on the curvature scalar or the behaviour at infinity, and are suitable for application in other problems, such as the Kerr solution and the no-hair theorems.  
These issues will be discussed in a future publication.
 
\section*{Appendix}

The tidal forces at the horizon in 
Schwarszchild's black hole are such that \cite{hob}
$$
\left.R^{\hat 1}_{\;\hat 0\hat 0 \hat 1}\right|_0>0,\;\;\;\;\;\;
\left.R^{\hat 2}_{\;\hat 0\hat 0 \hat 2}\right|_0<0 .
$$
In the geometry given by Eqn.(\ref{metric}), 
$$
\left.R^{\hat 1}_{\;\hat 0\hat 0 \hat 1}\right|_0 =\frac{b''_0r_0-(2+3\phi'_0r_0)(b'_0-1)}{2r_0^2},
$$
$$
\left.R^{\hat 2}_{\;\hat 0\hat 0 \hat 2}\right|_0 = \left.R^{\hat 3}_{\;\hat 0\hat 0 \hat 3}\right|_0 = \frac{b'_0-1}{2r_0^2}.
$$

\end{document}